\documentclass[reprint,aps,superscriptaddress,nofootinbib,]{revtex4}

\usepackage{amsmath}
\usepackage{graphicx}
\usepackage{color}
\usepackage{slashed}

\begin{document}

\title{Finiteness of Entanglement Entropy in Quantum Black Hole}
\author{Wen-Yu Wen}\thanks{%
E-mail: steve.wen@gmail.com}
\affiliation{Department of Physics and Center for High Energy Physics, Chung Yuan Christian University, Chung Li City, Taiwan}
\affiliation{Leung Center for Cosmology and Particle Astrophysics\\
National Taiwan University, Taipei 106, Taiwan}

\begin{abstract}
A logarithmic but divergent term usually appears in the computation of entanglement entropy circumferencing a black hole, while the leading quantum correction to the Bekenstein-Hawking entropy also takes the logarithmic form.  A quench model of CFT within finite Euclidean time was proposed in the \cite{Kuwakino:2014nra} to regard this logarithmic term as entanglement between radiation and the black hole, and this proposal was justified by the alternative sign for $n$-partite quantum information.  However, this preliminary form suffers from the notorious divergence at its low temperature limit.  In this letter, we propose a modified form for black hole entanglement entropy such that the divergence sickness can be cured.  We discuss the final stage of black hole due to this modification and its relation to R{\`e}nyi entropy, higher loop quantum correction and higher spin black holes.
\end{abstract}

\maketitle



\section{Introduction}

\begin{figure}[htbp]
  \begin{center}
    \includegraphics[clip,width=10.0cm]{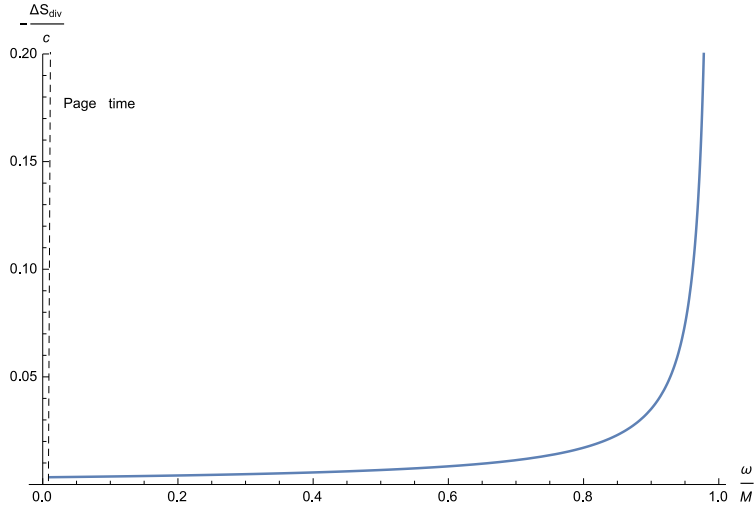}
    \caption{ \label{fig:Page_time}Proposed one-partite information per central charge grows exponentially at later stage of radiation.  The Page time is defined at the half of Bekenstein-Hawking entropy.  For illustrative purpose, we draw the Page time for $M/\omega \sim {\cal O}(500)$.  Practically a stellar black hole could weight $10^{40}$ in the Planck unit and Page time would be at $\omega/M \sim {\cal O}(10^{-40})$.}
   
  \end{center}
\end{figure}

We have learnt that a logarithmic divergence could occur in the computation of entropy  $\sigma$ for stationary scalar fields thermalized with infinite large Schwarzschild black hole in the near-horizon Rindler space\cite{Susskind:1994sm,Fiola:1994ir};
\begin{equation}
\sigma \sim \frac{1}{6} \ln \frac{\l}{\epsilon},
\end{equation}
for the IR cutoff scale $\l$ and  UV cutoff scale $\epsilon$.   We remark that the coefficient is related to the conformal anomaly and different for fermions and gauge fields.  In the case of gauge fields, to match the perpendicular component of electromagnetic field across the entangling surface, one has to include the edge modes in order to reproduce the universal result for log-divergent term \cite{Donnelly:2014fua,Huang:2014pfa}.
At the finite Hawking temperature, a similar time-independent logarithmic divergence appears in the computation of entanglement entropy from CFT side and can be verified in the holographic picture using BTZ black string\cite{Hartman:2013qma}
\begin{equation}\label{log_div}
S_{div} = \frac{c}{6}\ln (\frac{\beta}{4\pi\epsilon})
\end{equation}
In particular, $\beta=2\pi\l/\sqrt{M}$ for nonrotating BTZ black hole.  Regardless its divergence, one can obtain a finite quantity by comparing this logarithmic term before and after the black hole radiates a mass or energy quantum $\omega$.  We define the difference
\begin{equation}\label{log_diff}
\Delta S_{div} = S_{div}\big|_{M-\omega} - S_{div}\big|_{M} = -\frac{c}{12}\ln (1-\frac{\omega}{M}).
\end{equation}
It is tempted to think that this finite quantity associates with the information leakage via its radiation.   The one-partite information could be defined as the loss of entropy, that is $I^{[1]}=-\Delta S_{div}$\cite{Kuwakino:2014nra}.  It grows exponentially at later stage of evaporation, as shown in the figure \ref{fig:Page_time}.  It is amusing to see that retentive information is only available to observer outside the horizon at very late stage, as first pointed out in the \cite{Page:1993wv}.

The logarithmic form (\ref{log_div}) reminds us of a universal logarithmic correction to the Bekenstein-Hawking entropy:
\begin{equation}\label{log_correction}
S^{q}_{BH} = \alpha \ln \frac{A}{\l_p^2}
\end{equation}
In the case of BTZ black hole in AdS$_3$ with radius of curvature $\l$, where $A=2\pi \l \sqrt{M}$, equation (\ref{log_correction}) takes same form as (\ref{log_div}) if we recall $\beta= 8\pi \l/\sqrt{M}$, and the UV scale can be set by the Planck length, say $\epsilon \sim \l_{p}$.   The central charge $c$ encodes the degrees of freedom in the field theory, while the constant of proportionality $\alpha$ could depend on species and spins of radiated particles\cite{Fursaev:1994te}.  These similarities between (\ref{log_div}) and (\ref{log_correction}) as summarized in the Table \ref{table:similar}, have made us propose a $(0+1)-$dimensional quench model such that a massive radiated particle comes in sudden entangled with a black hole of mass $M$ would carry the entanglement entropy $S_E \sim \ln (M/m_{p})$ at late time\cite{Kuwakino:2014nra}.  In fact, the entropy difference (\ref{log_diff}) agrees with the definition of one-partite information in our quench model and the two-partite information are closely related to the definition of mutual information in the Parikh-Wilczek tunneling model\cite{Parikh:1999mf,Zhang:2009jn,Chen:2009ut}.  In addition, the $n$-partite quantum information, which carries alternative signs for even or odd $n$, also agrees with that computed in the holographic model\cite{Alishahiha:2014jxa}.  We summarize these results in the Appendix A.

In this letter, we would like to further investigate this similarity between entanglement entropy in the conformal field theory and that in the black hole radiation.  To be specific, while the seemly finite one-partite information (\ref{log_diff}) could still suffer from divergence when the black hole becomes Planck size, say $M\to \omega$.  We would like to point out that this potential sickness in the black hole entanglement entropy could be cured in a similar manner as that in the field theory side.  We discuss its impact to the final stage of black hole, and its relation to higher loop correction and higher spin to the Hawking temperature.  This paper is organized as follows: In the section II, using the replica trick, we propose a modified form for (\ref{log_correction}) such that the entanglement entropy remains finite at its IR limit.  In the section III,  we discuss the impact of modified entanglement entropy to its IR behavior and show its connection to the higher loop quantum correction and higher spin through modification of the Hawking temperature.  At last, we briefly introduce the definition of n-partite information from quantum quench model and derive the corrected Hawking temperature in higher spin black holes in the Appendices. 

\begin{table}
\begin{center}
    \begin{tabular}{ c | c | c }
     & CFT & BTZ  \\ \hline
    time scale & time $t$ & inverse Hawking temperature $\beta$ \\ \hline
    UV scale &  cutoff length $\epsilon$ & Planck length $\l_p$ \\ \hline
    degrees of freedom & central charge $c$ & $1-$loop coefficient $\alpha_1$ \\ \hline
    entanglement entropy &  $c\ln \frac{t}{\epsilon}$ & $\alpha_1\ln \frac{\beta}{\l_p}$ \\
\hline
    \end{tabular}
\end{center}
\caption{\label{table:similar}Similarities between late time entanglement entropy in CFT and proposed logarithmic entanglement entropy of BTZ}
\end{table}

\section{BTZ black hole and finite entanglement entropy}
For simplicity, we will focus on the nonrotating BTZ black hole in the asymptotic $AdS$ with curvature radius $\l$.  The metric reads
\begin{equation}
-(-M+\frac{r^2}{\l^2})dt^2+(-M+\frac{r^2}{\l^2})^{-1}dr^2+r^2 d\phi^2,
\end{equation}
where the horizon is given by $r_+=\l \sqrt{M}$.  The Hawking temperature and Bekenstein-Hawking entropy are given by  
\begin{eqnarray}
&&T_{H}=\beta^{-1}=\frac{\sqrt{M}}{2\pi \l},\\
\label{eqn:BTZ_entropy}&&S_{BH} = \frac{A_H}{4G}=\frac{\pi \l \sqrt{M}}{2G},\\
\end{eqnarray}

\begin{figure}[htbp]
  \begin{center}
    \includegraphics[clip,width=10.0cm]{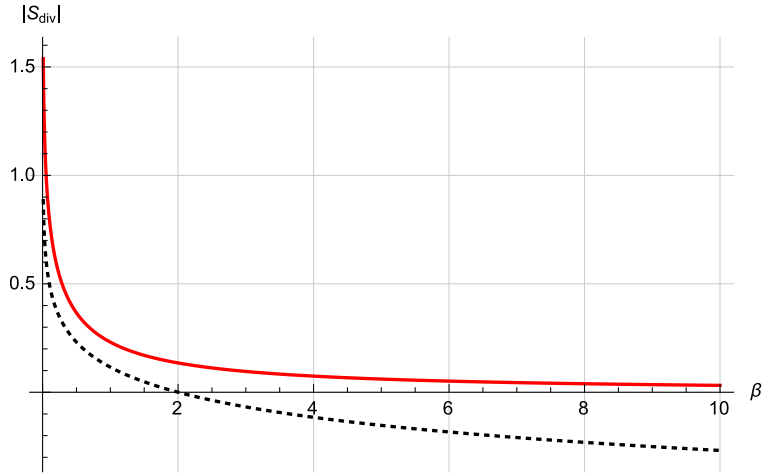}
    \caption{Divergent logarithmic growth with $\beta$ (\ref{log_div}) (dotted black) and its regularized form (\ref{finite_entanglement}) (solid red).  The regularized logarithmic term approaches constant at large $\beta$ or low temperature. }
    \label{fig:log_entanglement}
  \end{center}
\end{figure}

Before a regularized form for (\ref{log_diff}) could be proposed, we would like to obtain some insights from the entanglement entropy in the CFT.  In the analysis of $2m$-point functions, the growth of entanglement entropy $\Delta S_A^{(m)}$ is suggested to take following form\cite{Caputa:2014vaa}:
\begin{equation}
\Delta S_A^{(m)} \simeq -\frac{1}{m-1}\ln \big[ \frac{1}{D_m} + \mu_m \cdot (\frac{\epsilon}{t})^{\nu_m} \big], 
\end{equation}
for positive $D_m$ and $\nu_m$.   The entanglement entropy grows like log$(t)$ at small $t$, but approaches a finite constant for large $t$.  The above formulas can be obtained by replica trick\cite{CC1}, where the regularized R{\`e}nyi entropy is defined as 
\begin{equation}
I_n=\big[ \frac{1}{D_n}+ \mu_n \cdot (\frac{\epsilon}{t})^{\nu_n}\big]^{\Delta_n},
\end{equation}
for $\Delta_n = \frac{c}{12}(n-\frac{1}{n})$.  The two-point function or entanglement entropy is obtained by taking the limit
\begin{equation}
S_A = -\partial_n I_n \big|_{n=1} = -\frac{c}{3}\ln \big[ \frac{1}{D_1} + \mu_1 \cdot (\frac{\epsilon}{t})^{\nu_1} \big].
\end{equation}
Similarly, at the finite temperature, we propose the R{\`e}nyi entropy, originally defined in \cite{Hartman:2013qma}, would be regularized as 
\begin{equation}
I_n = \big[ \frac{1}{D_n}+ \frac{2\pi\epsilon}{\beta} sech(\frac{2\pi t}{\beta}) \big]^{\Delta_n},
\end{equation} 
which gives rise to the regularized entanglement entropy
\begin{equation}
S_A = -\frac{c}{3}\ln \big[ \frac{1}{D_1} + \frac{2\pi\epsilon}{\beta} sech(\frac{2\pi t}{\beta})\big]
\end{equation}
The time-independent {\sl divergent} piece can be retrieved at limit $t\to 0$, that is
\begin{equation}\label{finite_entanglement}
S_{div} ^\prime= -\frac{c}{3} \ln \big[ \frac{1}{D_1}+\frac{2\pi\epsilon}{\beta} \big].
\end{equation}
We remark that this regularized term is now finite at the limit $\epsilon\to 0$ or $\beta \to \infty$.  The regularized one-partite information $I^{[1]}=-\Delta S_{div} ^\prime$ also becomes finite at limit $M\to \omega$.  We compare the behavior of (\ref{log_div}) and (\ref{finite_entanglement}) in the Fig. \ref{fig:log_entanglement}. 

\section{Perturbative correction to Hawking temperature}

Taking the perturbative approach towards quantum black holes, the Hawking temperature is expected to receive one-loop quantum correction as follows\cite{Govindarajan:2001ee,Banerjee:2008ry}:
\begin{equation}
T^{[1]}_H = T_H/ (1+\alpha_1\frac{\hbar}{r_+}).
\end{equation}
As a result, the Bekenstein-Hawking entropy is corrected by a logarithmic term\cite{Fursaev:1994te,Carlip:2000nv}:
\begin{equation}\label{BTZ_entropy}
S^{[1]}_{BH} = \int{\frac{dM}{T^{[1]}_{H}}}=\frac{\pi r_+}{2G} + \alpha_1\hbar\frac{\pi}{2G}\ln r_+ 
\end{equation}
To avoid divergence, the second logarithmic term is expect to be regularized at the limit $r_+\to 0$.   If one assumes the logarithmic correction term behaves just like the entanglement entropy as discussed in previous section, one may guess the regularized form
\begin{equation}\label{BTZ_entropy_regular}
S^{[1]\prime}_{BH} = \frac{\pi r_+}{2G}   -c \ln\big[ \frac{1}{D} +\frac{2\pi\l_p}{\beta}\big],
\end{equation}
for  a positive constant $c=\frac{|\alpha_1|\pi\hbar}{2G}$ given $\alpha_1<0$\cite{Carlip:2000nv}.   Here we have explicitly used Euclidean time period $\beta$ and chose the Planck length $\l_p$ as UV cutoff scale.  We remark that the IR divergence of (\ref{BTZ_entropy}) is in fact consistent with discussion in the \cite{Mallayya:2014xwa} where it was argued that the IR divergence is due to the scaling symmetry in the free two-dimensional CFT.  However, we expect a finite system can only store finite amount of entanglement entropy and this calls for some boundary effect or nonperturbative interaction, which gives rise to the regularized form in (\ref{BTZ_entropy_regular}).

We would like to comment on the physical meaning of $D$.  If the BTZ completely evaporates at final stage with zero temperature, then (\ref{BTZ_entropy_regular}) leads to a nonzero value $c\ln D$.  Since there is nothing left to host this entropy, it is better to set $D=1$.  On the other hand, we have learnt that $D_n$ is related to the quantum dimension $d_{\cal O}$ through $D_n = (d_{\cal O})^{n-1}$ \cite{He:2014mwa}.  If $D$ were the very $D_1$ in the (\ref{finite_entanglement}), then it also suggests that $D=(d_{\cal O})^0=1$.  This may imply that all the degrees of freedom responsible to entanglement are frozen in the ground state.

With a choice of $\alpha_1 = \frac{\l^2}{\hbar\l_p}$, the correction term, which was argued to be related to the entanglement entropy, has the same form as the R{\`e}nyi entropy:
\begin{equation}
S_{R} = \frac{1}{c}\ln\big[1+ cS_{BH}\big],
\end{equation}
for $c=\frac{\pi \l^2}{2G\l_p}$ and BTZ entropy $S_{BH}$ given in (\ref{eqn:BTZ_entropy}).

\begin{figure}[htbp]
  \begin{center}
    \includegraphics[clip,width=10.0cm]{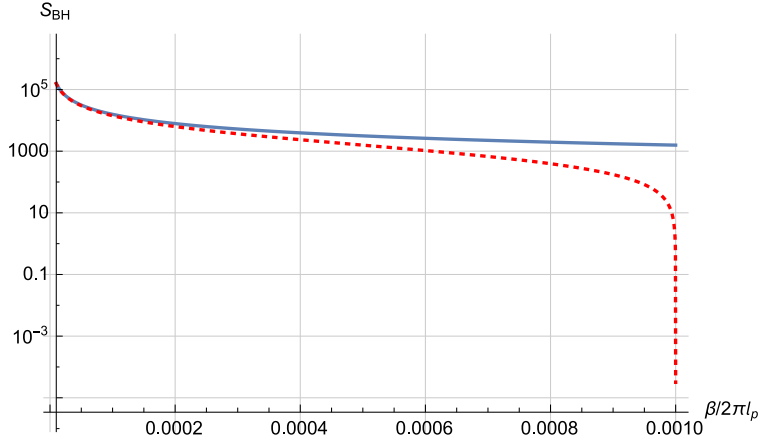}
    \caption{Logarithmic corrected Bekenstein-Hawking entropy versus $\beta$ for different choices of $D$.  At the final stage, the black hole could evaporate completely (red dotted) or a warm remnant forms (blue solid).}
    \label{fig:BH_entropy}
  \end{center}
\end{figure}

Another scenario is to assume there is a warm remnant at the final stage\cite{Adler:2001vs,Chen:2014jwq,Xiang:2006mg}, say $\beta = 2\pi \l_p$, then we would have
\begin{equation}
D^{-1} = \exp \big[\frac{\pi \l^2}{2cG\l_p}-\frac{S_{rem}}{c} \big]-1,
\end{equation}
where the remnant carries entropy S$_{rem}$.  We plot the black hole entropy versus $\beta$ in the figure (\ref{fig:BH_entropy}).

The value of $D$ can also be determined if one further considers two-loop quantum correction to the Hawking temperature:
\begin{equation}\label{2-loop}
T^{[2]}_H = T_H/ (1+\alpha_1\frac{\hbar}{r_+} + \alpha_2\frac{\hbar^2}{r_+^2})
\end{equation} 
such that
\begin{equation}\label{entropy_2-loop}
S^{[2]}_{BH} = \int{\frac{dM}{T^{[2]}_{H}}}=\frac{\pi r_+}{2G} + \alpha_1\hbar\frac{\pi}{2G}\ln r_+ - \alpha_2 \hbar^2 \frac{\pi}{2G r_+},
\end{equation}
Now we expand $\beta$ in the (\ref{BTZ_entropy_regular}) around $2\pi \l_p$:
\begin{equation}\label{UV_expansion}
S^{[1]\prime}_{BH} \simeq \frac{\pi r_+}{2G}  -c \ln \frac{2\pi \l_p}{\beta} -c \ln (1+\frac{\beta}{D}) \simeq \Lambda -c \ln \frac{2\pi\l_p}{\beta} -c \frac{\beta}{4\pi\l_p D} + \cdots.
\end{equation}
The additional linear $\beta$ term corresponds to last term in  (\ref{entropy_2-loop}) and one obtains
\begin{equation} 
D = \frac{\l^2|\alpha_1|}{2\l_p\hbar \alpha_2}. 
\end{equation}

In conclusion, both choices of $D$ have suggested that (\ref{BTZ_entropy_regular}) would be a good academic guess for a regularized black hole entropy.  At least it either serves as an effective model at final stage of evaporation, or provides a close form to match up with two-loop quantum correction.

We have discussed the connection between large radius expansion of (\ref{BTZ_entropy_regular}) and loop corrections (\ref{entropy_2-loop}).  Now we would like to explore its possible connection to the higher spin gravity.   Tensor fields of higher spin are easily found in the higher dimensional theory such as superstring.  The entanglement corresponds to the conical entropy with surface term contribution and the string scale serves as a natural cutoff\cite{He:2014gva}.  The higher spin black holes in $AdS_3$ were studied in the Chern-Simons formalism \cite{Gutperle:2011kf,Kraus:2011ds} and the leading correction to the Hawking temperature, which is derived in the appendix, has the following form:
\begin{equation}\label{Hawking_T_highspin}
T_H^{[3]} \simeq T_H/(1- \frac{16G|\sigma| \mu^2}{\l}r_+^2+\cdots),
\end{equation}
where $\sigma<0$ is the arbitrary constant used to normalize the high spin generator and $\mu$ is the source or coupling to the high spin current.  According to the first law of thermodynamics, this leads to a correction to the thermal entropy
\begin{equation}
S_{BH}^{[3]} = \frac{\pi r_+}{2G} - \frac{8\pi |\sigma| \mu^2}{3\l^3}r_+^3+\cdots.
\end{equation}
In comparison with the small radius expansion of (\ref{BTZ_entropy_regular}), which reads
\begin{equation}
S_{BH}^{[1]\prime}=\frac{\pi r_+}{2G} + c\ln{D} -\frac{2\pi \l c D}{\beta} +  \frac{2\pi^2\l^2 c D^2}{\beta^2}   -\frac{8\pi^3 \l^3 c D^3}{3\beta^3} + \cdots.
\end{equation}
We see the higher spin correction may contribute to order ${\cal O}(\beta^{-3})$ and after.

As a final remark, in this letter we only discuss the black hole whose inverse Hawking temperature $\beta$ only depends on single charge $M$.  Therefore one can easily establish connection between real time $t$ and black hole mass $M$.  It would be interesting to further investigate the applicability of similar entanglement model to those black holes with multiple charges.

\begin{acknowledgments}
The author is grateful to inspiring talks and discussion during the KEK theory workshop at the late stage of this project.   This work is supported in parts by the Taiwan's Ministry of Science and Technology (grant No. 102-2112-M-033-003-MY4 and No. 103-2633-M-033-003-). 
\end{acknowledgments}

\appendix

\section{n-partite information from quantum quench model}

In \cite{Kuwakino:2014nra}, we have proposed a quench model stating that a quantum state comes in sudden entangled with a black hole would carry the entanglement entropy $S_E \sim \ln (M/m_{p})$ at later time and this entanglement is responsible for logarithmic correction to the Bekenstein-Hawking entropy.  Then it is suggested to define the one-partite information as {\sl relative} entropy between two different black hole masses: the original black hole mass and that after emitting a particle with mass $\omega_i$.  That is,
\begin{equation}
I_i^{[1]} = 8\pi \alpha (\ln (M-\omega_i)/m_{p} - \ln M/m_{p}) =8\pi\alpha \ln (1-\frac{\omega_i}{M}).
\end{equation}
Then the bipartite information among two emitted quanta $\omega_i$ and $\omega_j$ can be defined by the conventional way:
\begin{equation}\label{eqn:bipartite_quantum}
I^{[2]}_{ij}\equiv I^{[1]}_{i} + I^{[1]}_{j} - I^{[1]}_{(ij)} =  8\pi \alpha \ln{\frac{(M-\omega_i)(M-\omega_j)}{M(M-\omega_i-\omega_j)}}.
\end{equation}
This bipartite information exactly captures the quantum part in the definition of {\sl mutual information}, which measures how does the entropy change for an emission $\omega_j$ depend on whether an earlier emission $\omega_i$ happens or not, that is 
\begin{equation}\label{eqn:mutualinfo}
E^{[2]}_{ij} \equiv S_E(M,\omega_j \big|\omega_i)-S_E(M,\omega_j)= 8\pi \omega_i\omega_j + I^{[2]}_{ij},
\end{equation}
where $S_E(M^\prime,\omega_i) \equiv S_{BH}\big|_{M^\prime=M} - S_{BH}\big|_{M^\prime=M-\omega_i}$ and $S_E(M,\omega_j \big|\omega_i) = S_E(M-\omega_i,\omega_j)$ for the tunneling model \cite{Parikh:1999mf,Zhang:2009jn}.  After all, the $n$-partite information can be computed iteratively from $(n-1)$-partite:
\begin{eqnarray}
I^{[n]}_{i_1i_2\cdots i_{n-1}} &&= I^{[n-1]}_{i_1i_2\cdots i_{n-2}i_{n-1}}+I^{[n-1]}_{i_1i_2\cdots i_{n-2}i_{n}}-I^{[n-1]}_{i_1i_2\cdots (i_{n-1} i_{n})}\nonumber\\
&&=8\pi\alpha \ln{\frac{ {\displaystyle \prod_{p} } (M-\omega_{i_p}) { \displaystyle \prod_{p<s<t} }(M-\omega_{i_p}-\omega_{i_s}-\omega_{i_t})\cdots}{M { \displaystyle \prod_{p<s} }(M-\omega_{i_p}-\omega_{i_s}) { \displaystyle \prod_{p<s<t<u} } (M-\omega_{i_p}-\omega_{i_s}-\omega_{i_t}-\omega_{i_u})\cdots}},
\end{eqnarray}
where indexes $p, s, t, u, \cdots$ run through $1, 2, \cdots, n$.  It has been shown that the $n$-partite information is always positive for even $n$ and negative for odd $n$\cite{Kuwakino:2014nra}.

\section{Higher spin correction to the Hawking temperature}

We would like to show that in the small radius or small mass limit, one can regard the higher spin contribution as a correction to the Hawking temperature as shown in (\ref{Hawking_T_highspin}).  We start with the conclusion in the \cite{Gutperle:2011kf} stating that the $AdS_3$ black hole with spin-$3$ charge has a linearized solution and its entropy can be casted in the following form:
\begin{equation}
S({\cal L},{\cal W}) = 2\pi \sqrt{2\pi k {\cal L}} f(y), \quad y = \frac{27k{\cal W}^2}{-64 \sigma \pi {\cal L}^3},
\end{equation}
where $k=\l/4G$, ${\cal L}$ and ${\cal W}$ are the stress tensor and spin-$3$ current.  $f(y)$ is a function between $0$ and $2$.  The linearized smooth and holonomy condition sets ${\cal W}=-\frac{-64\pi\sigma \mu}{3k}{\cal L}^2$.  For the nonrotating BTZ and small $\mu$, one has $f(y) \simeq 1 -\frac{1}{16}y +\cdots$ and the leading order correction to $\beta$ can be calculated as 
\begin{equation}
\beta = \frac{1}{2\pi}\frac{\partial S}{\partial {\cal L}} \simeq \beta_0 (1-\frac{4|\sigma|\mu^2}{k}M),
\end{equation}
where ${\cal L}=\frac{M}{4\pi}$ and $\beta_0=\sqrt{\frac{\pi k}{2{\cal L}}}$.  Similar result can also be obtained if one starts with higher spin partition function as a series expansion of $\frac{\mu}{\tau}$ \cite{Kraus:2011ds}.


\end{document}